\documentclass[11pt]{article}
\usepackage{graphicx}
\usepackage{amsmath}

\newcommand{\BABARPubYear}    {06}

\newcommand{\BABARConfNumber} {16}
\newcommand{\SLACPubNumber} {12012}

\RequirePackage{xspace}

\usepackage{relsize}
\def\babar{\mbox{\slshape B\kern-0.1em{\smaller A}\kern-0.1em
    B\kern-0.1em{\smaller A\kern-0.2em R}}}

\def\epem       {\ensuremath{e^+e^-}\xspace}

\def\q     {\ensuremath{q}\xspace}

\def\qqbar {\ensuremath{q\overline q}\xspace}
\def\u     {\ensuremath{u}\xspace}
\def\ubar  {\ensuremath{\overline u}\xspace}

\def\d     {\ensuremath{d}\xspace}

\def\s     {\ensuremath{s}\xspace}

\def\c     {\ensuremath{c}\xspace}

\def\b     {\ensuremath{b}\xspace}

\def\piz   {\ensuremath{\pi^0}\xspace}

\def\kaon  {\ensuremath{K}\xspace}
\def\Kbar  {\kern 0.2em\overline{\kern -0.2em K}{}\xspace}

\def\Kz    {\ensuremath{K^0}\xspace}
\def\Kzb   {\ensuremath{\Kbar^0}\xspace}
\def\KzKzb {\ensuremath{\Kz \kern -0.16em \Kzb}\xspace}
\def\Kp    {\ensuremath{K^+}\xspace}
\def\Km    {\ensuremath{K^-}\xspace}

\def\KpKm  {\ensuremath{\Kp \kern -0.16em \Km}\xspace}

\def\Dbar    {\kern 0.2em\overline{\kern -0.2em D}{}\xspace}

\def\Dz      {\ensuremath{D^0}\xspace}
\def\Dzb     {\ensuremath{\Dbar^0}\xspace}
\def\DzDzb   {\ensuremath{\Dz {\kern -0.16em \Dzb}}\xspace}
\def\Dp      {\ensuremath{D^+}\xspace}
\def\Dm      {\ensuremath{D^-}\xspace}

\def\DpDm    {\ensuremath{\Dp {\kern -0.16em \Dm}}\xspace}

\def\B       {\ensuremath{B}\xspace}
\def\Bbar    {\kern 0.18em\overline{\kern -0.18em B}{}\xspace}

\def\BB      {\ensuremath{B\Bbar}\xspace} 
\def\Bz      {\ensuremath{B^0}\xspace}
\def\Bzb     {\ensuremath{\Bbar^0}\xspace}
\def\BzBzb   {\ensuremath{\Bz {\kern -0.16em \Bzb}}\xspace}
\def\Bu      {\ensuremath{B^+}\xspace}
\def\Bub     {\ensuremath{B^-}\xspace}
\def\Bp      {\ensuremath{\Bu}\xspace}

\def\Bpm     {\ensuremath{B^\pm}\xspace}

\def\BpBm    {\ensuremath{\Bu {\kern -0.16em \Bub}}\xspace}

\def\BorBbar    {\kern 0.18em\optbar{\kern -0.18em B}{}\xspace}
\def\DorDbar    {\kern 0.18em\optbar{\kern -0.18em D}{}\xspace}
\def\KorKbar    {\kern 0.18em\optbar{\kern -0.18em K}{}\xspace}

\mathchardef\Upsilon="7107
\def\Y#1S{\ensuremath{\Upsilon{(#1S)}}\xspace}

\def\FourS {\Y4S}

\mathchardef\Deltares="7101
\mathchardef\Xi="7104
\mathchardef\Lambda="7103
\mathchardef\Sigma="7106
\mathchardef\Omega="710A

\def\Deltabar{\kern 0.25em\overline{\kern -0.25em \Deltares}{}\xspace}
\def\Lbar{\kern 0.2em\overline{\kern -0.2em\Lambda\kern 0.05em}\kern-0.05em{}\xspace}
\def\Sigbar{\kern 0.2em\overline{\kern -0.2em \Sigma}{}\xspace}
\def\Xibar{\kern 0.2em\overline{\kern -0.2em \Xi}{}\xspace}
\def\Obar{\kern 0.2em\overline{\kern -0.2em \Omega}{}\xspace}
\def\Nbar{\kern 0.2em\overline{\kern -0.2em N}{}\xspace}
\def\Xb{\kern 0.2em\overline{\kern -0.2em X}{}\xspace}

\def\upsbb   {\ensuremath{\FourS \to \BB}\xspace}

\def\pt         {\mbox{$p_T$}\xspace}
\def\mes        {\mbox{$m_{\rm ES}$}\xspace}

\def\DeltaE     {\mbox{$\Delta E$}\xspace}

\newcommand{\tev}{\ensuremath{\mathrm{\,Te\kern -0.1em V}}\xspace}
\newcommand{\gev}{\ensuremath{\mathrm{\,Ge\kern -0.1em V}}\xspace}
\newcommand{\mev}{\ensuremath{\mathrm{\,Me\kern -0.1em V}}\xspace}
\newcommand{\kev}{\ensuremath{\mathrm{\,ke\kern -0.1em V}}\xspace}
\newcommand{\ev}{\ensuremath{\mathrm{\,e\kern -0.1em V}}\xspace}
\newcommand{\gevc}{\ensuremath{{\mathrm{\,Ge\kern -0.1em V\!/}c}}\xspace}
\newcommand{\mevc}{\ensuremath{{\mathrm{\,Me\kern -0.1em V\!/}c}}\xspace}
\newcommand{\gevcc}{\ensuremath{{\mathrm{\,Ge\kern -0.1em V\!/}c^2}}\xspace}
\newcommand{\mevcc}{\ensuremath{{\mathrm{\,Me\kern -0.1em V\!/}c^2}}\xspace}

\def\mus  {\ensuremath{\rm \,\mus}\xspace}

\def\ps   {\ensuremath{\rm \,ps}\xspace}

\def\mus        {\ensuremath{\,\mu{\rm s}}\xspace}    
\def\ps         {\ensuremath{{\rm \,ps}}\xspace}  

\def\to                 {\ensuremath{\rightarrow}\xspace}

\newcommand{\stat}{\ensuremath{\mathrm{(stat)}}\xspace}
\newcommand{\syst}{\ensuremath{\mathrm{(syst)}}\xspace}

\def\pep2{PEP-II}

\def\gsim{{~\raise.15em\hbox{$>$}\kern-.85em
          \lower.35em\hbox{$\sim$}~}\xspace}
\def\lsim{{~\raise.15em\hbox{$<$}\kern-.85em
          \lower.35em\hbox{$\sim$}~}\xspace}

\def\CP                {\ensuremath{C\!P}\xspace}
\def\C       {\ensuremath{C}\xspace}

\def\deltaz{\ensuremath{{\rm \Delta}z}\xspace}
\def\deltat{\ensuremath{{\rm \Delta}t}\xspace}
\def\deltamd{\ensuremath{{\rm \Delta}m_d}\xspace}

\xspace

\newcommand{\epjBase}        {Eur.\ Phys.\ Jour.\xspace}
\newcommand{\jprlBase}       {Phys.\ Rev.\ Lett.\xspace}
\newcommand{\jprBase}        {Phys.\ Rev.\xspace}
\newcommand{\jplBase}        {Phys.\ Lett.\xspace}
\newcommand{\nimBaseA}       {Nucl.\ Instr.\ Methods Phys.\ Res., Sect.\ A\xspace}

\newcommand{\epjc}      [1]  {\epjBase\ C~{\bf #1}}

\newcommand{\mpl}       [1]  {{Mod.\ Phys.\ Lett.\ {\bf #1}}}

\newcommand{\nima}      [1]  {\nimBaseA~{\bf #1}}

\newcommand{\plb}       [1]  {\jplBase\ B~{\bf #1}}

\newcommand{\jprl}      [1]  {\jprlBase\ {\bf #1}}
\newcommand{\jprd}      [1]  {\jprBase\ D~{\bf #1}}

\newcommand{\progtp}    [1]  {{Prog.\ Theor.\ Phys.\ {\bf #1}}}

\def\jetset74   {\mbox{\tt Jetset \hspace{-0.5em}7.\hspace{-0.2em}4}\xspace}

\newcommand{\su}     [1]  {\ensuremath{SU(#1)}}

\newcommand{\e}      [1]   { {\ensuremath{ \times 10^{ {#1} } }}}

\def\Bch   {\ensuremath{B^{\pm}}}

\def\mes   {\ensuremath{m_{ES}}}
\def\de   {\ensuremath{\Delta E}}
\def\dt   {\ensuremath{\Delta t}}

\def\coshel {\ensuremath{ cos( \theta_{H}) }}
\def\mv   {\ensuremath{ M_V }}

\def\dm   {\ensuremath{\Delta m}}

\def\pt   {\ensuremath{ p_t }}

\def\rhoz   {\ensuremath{\rho^{0}}}

\def\bb   {\ensuremath{B \overline{B}}}
\def\ifb  {\ensuremath{fb^{-1}}}

\def\SCF   {{\ensuremath{\mathrm{SCF}}}}
\def\true  {{\ensuremath{\mathrm{true}}}}
\def\Long  {{\ensuremath{\mathrm{long}}}}
\def\Tran  {{\ensuremath{\mathrm{tran}}}}

\def\clong   {\ensuremath{C_{\Long}}}
\def\slong   {\ensuremath{S_{\Long}}}
\def\ctran   {\ensuremath{C_{\Tran}}}
\def\stran   {\ensuremath{S_{\Tran}}}

\def\ptrue   {\ensuremath{f_{L}}}

\def\aeff {\ensuremath{\alpha^{\rho\rho}_{eff}}}

\def\piz {\ensuremath{\pi^0}}

\def\de       {\ensuremath{\Delta E}}

\def\borhorho {\ensuremath{B^{0} \rightarrow \rho^+ \rho^- }}

\def\Bztorhoprhom {\ensuremath{\Bz (\Bzb) \to \rho^+ \rho^- }\xspace}

\def\rhop {\ensuremath{\rho^+ }\xspace}

\def\coshel  {\ensuremath{ \cos\theta_{i} }}
\def\mv      {\ensuremath{ m_{\pi^\pm \pi^0 }}}

\def\nno     {{\ensuremath{\cal{N}}}}

\def\CPV {\ensuremath{CPV}}

\def\Bztorhozrhoz {\ensuremath{\,\Bz \to \rho^0\rho^0}}
\def\Bztorhoprhom {\ensuremath{\,\Bz \to \rho^+\rho^-}}

\def\lepton   {{\tt Lepton} }
\def\kaon     {{\tt Kaon} }
\def\kaonpion {{\tt Kaon-Pion} }
\def\pion     {{\tt Pion} }
\def\other    {{\tt Other} }
\def\notag    {{\tt Untagged} }

\def\lumi {\ensuremath{316\, \ifb}}
\def\nbb {\ensuremath{(347\pm 3.9)\e{6}\, \bb\ \rm pairs}}
\def\offpeaklumi {\ensuremath{27.2\, \ifb}}
\def\ndata {33902} 
\def\correctedslong      {\ensuremath{-0.19 \pm 0.21 \stat}}
\def\correctedclong      {\ensuremath{-0.07 \pm 0.15 \stat}}
\def\correctedfl         {\ensuremath{0.977 \pm 0.024 \stat}}
\def\correctedsignalyield{\ensuremath{615 \pm 57 \stat}}

\def\sccorrelation       {-0.058}
\def\measuredslong {\ensuremath{\correctedslong ^{+0.05}_{-0.07} \syst}}
\def\measuredclong {\ensuremath{\correctedclong \pm 0.06 \syst}}
\def\measuredfl    {\ensuremath{\correctedfl ^{+0.015}_{-0.013} \syst}}
\def\measuredbf    {\ensuremath{(23.5 \pm 2.2 \stat \pm 4.1 \syst)\e{-6}}}
\def\measuredalpha {\ensuremath{[74, 117]^\circ}\, {\rm at}\, 68\%\, {\rm CL}}

\def\measureddeltaalpha{\ensuremath{18^\circ}}

\long\def\inst#1{\par\nobreak\kern 4pt\nobreak
    {\it #1}\par\vskip 10pt plus 3pt minus 3pt}

\begin{document}
{\pagestyle{empty}

\begin{flushright}
\babar-CONF-\BABARPubYear/\BABARConfNumber \\
SLAC-PUB-\SLACPubNumber \\
\end{flushright}

\par\vskip 1cm

\begin{center}
{\Large \bf Updated Measurement of the CKM Angle {\boldmath $\alpha$} Using {\boldmath \Bztorhoprhom} Decays}
\end{center}
\bigskip

\begin{center}
\large The \babar\ Collaboration\\
\mbox{ }\\
\today \\
\end{center}
\bigskip \bigskip

\begin{center}
\large \bf Abstract
\end{center}
We present results from an analysis of \Bztorhoprhom\ using 
\lumi\ of \upsbb\ decays observed with the \babar\ detector at the
\pep2\ asymmetric-energy $B$ Factory at SLAC. 
We measure the \Bztorhoprhom\ branching fraction, longitudinal polarization fraction \ptrue, and the \CP-violating parameters \slong\ and \clong:
\begin{eqnarray}
{\cal B}(\Bztorhoprhom) &=& \measuredbf,\nonumber \\
\ptrue &=& \measuredfl,\nonumber \\
\slong &=& \measuredslong,\nonumber \\
\clong &=& \measuredclong.\nonumber 
\end{eqnarray}
Using an isospin analysis of $B\rightarrow \rho\rho$ decays we determine the 
angle $\alpha$ of the unitarity triangle. 
One of the two solutions, $\alpha = \measuredalpha$,
is compatible with the standard model. All results presented here are preliminary.
\vfill
\begin{center}

Submitted to the 33$^{\rm rd}$ International Conference on High-Energy Physics, ICHEP 06,\\
26 July---2 August 2006, Moscow, Russia.

\end{center}

\vspace{1.0cm}
\begin{center}
{\em Stanford Linear Accelerator Center, Stanford University, 
Stanford, CA 94309} \\ \vspace{0.1cm}\hrule\vspace{0.1cm}
Work supported in part by Department of Energy contract DE-AC03-76SF00515.
\end{center}

\newpage
} 

%
%
\begin{center}
\small

The \babar\ Collaboration,
\bigskip

%
{B.~Aubert,}
{R.~Barate,}
{M.~Bona,}
{D.~Boutigny,}
{F.~Couderc,}
{Y.~Karyotakis,}
{J.~P.~Lees,}
{V.~Poireau,}
{V.~Tisserand,}
{A.~Zghiche}
\inst{Laboratoire de Physique des Particules, IN2P3/CNRS et Universit\'e de Savoie,
 F-74941 Annecy-Le-Vieux, France }
% ...

\end{center}\newpage

\section{INTRODUCTION}
\label{sec:Introduction}
\begin{figure}[!ht]
\begin{center}
\resizebox{6cm}{!}{\includegraphics{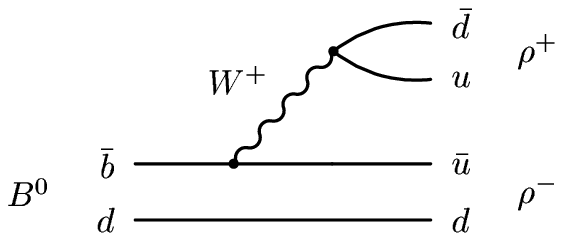}}
\resizebox{6cm}{!}{\includegraphics{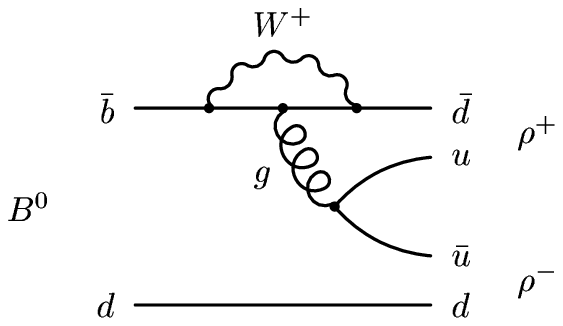}}
\caption{Tree and gluonic penguin diagrams contributing to the process \borhorho. The
         penguin contribution coming from the diagram with a top quark in the loop dominates
         as contributions from processes with $u$ and $c$ quarks are suppressed.}
\label{fig:feynmangraphs}
\end{center}
\end{figure}

In the standard model, charge conjugation-parity (\CP) violating effects in the \B-meson 
system arise from a single phase in the Cabibbo-Kobayashi-Maskawa 
(CKM) quark-mixing matrix~\cite{CKM}. Interference between the direct decay and
decay after $\Bz\Bzb$ mixing in $\Bztorhoprhom, \rho^\pm\pi^\mp, \pi^+\pi^-$ results 
in a time-dependent decay-rate asymmetry that is sensitive 
to the angle $\alpha \equiv \arg\left[-V_{td}^{}V_{tb}^{*}/V_{ud}^{}V_{ub}^{*}\right]$ 
in the unitarity triangle of the CKM matrix.
These decays mainly proceed through a $\b \to \u\ubar \d$ tree diagram.
The presence of penguin loop contributions introduces additional
phases that shift the experimentally measurable parameter
$\alpha_{\mathrm{eff}}$ away from the value of $\alpha$.
Figure~\ref{fig:feynmangraphs} shows the leading order tree and gluonic 
penguin loop contributions to this decay.  
Measurements of the $\Bp \to \rho^+\rho^0$~\cite{rhorho0} and $\Bz\to \rho^0 \rho^0$~\cite{PRLrho0rho0} branching fractions 
show that the penguin contribution in $\B \to \rho \rho$ is smaller than 
the leading order tree diagram~\cite{PRLrho0rho0}.  Results
from $B\to\pi\pi$ decays~\cite{babar_btopipi,belle_btopipi} and
$\Bz\to\rho^\pm\pi^\mp$~\cite{babar_btorhopi,belle_btorhopi} are 
discussed elsewhere~\cite{alphareview}.

\begin{figure}[!ht]
\begin{center}
\resizebox{12cm}{!}{\includegraphics{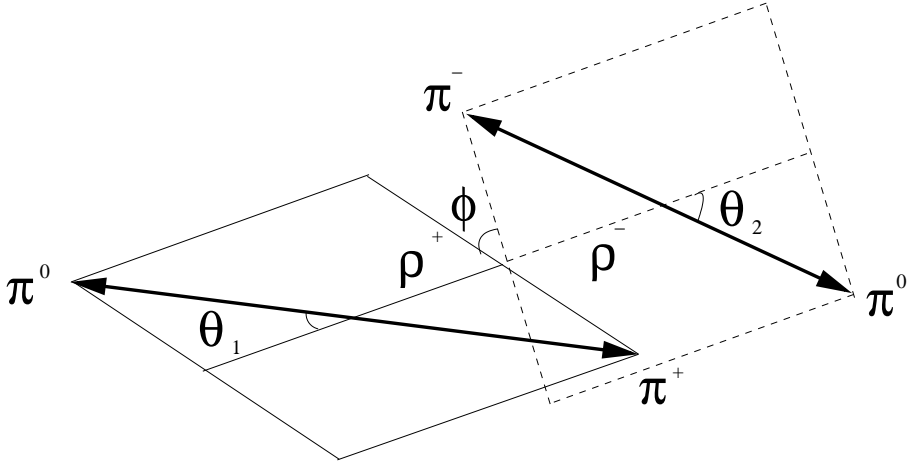}}
\caption{A schematic of the decay of a \B\ meson via two vector particles, 
         $\rho^+$ and $\rho^-$, into a four pion final state.  The $\rho$ meson
         final states are shown in their rest frames, and $\phi$ is the angle 
         between the decay planes of the $\rho$ mesons.}
\label{fig:btovv}
\end{center}
\end{figure}

In $\B\to\rho\rho$ decays, a spin 0 particle decays into two spin 1 
particles (as shown in Fig.~\ref{fig:btovv}).  Subsequently each 
$\rho$ meson decays into a $\pi\pi$ pair: $\rho^\pm\to\pi^\pm\pi^0$ 
and $\rho^0\to\pi^+\pi^-$.  As a result, the \CP analysis of $\Bz\to \rho^+ \rho^-$ is complicated 
by the presence of one mode with longitudinal polarization and 
two modes with transverse polarization.  The longitudinal mode is \CP 
even, while the transverse modes contain \CP-even and \CP-odd states.
The decay is observed to be dominated by the longitudinal 
polarization \cite{ref:us}, with a fraction $\ptrue$ defined as the 
fraction of the helicity zero state in the decay. Integrating over the 
angle between the 
$\rho$ decay planes $\phi$, the angular decay rate is
\begin{eqnarray}
\label{eqn:one} \frac{d^2\Gamma}{\Gamma d\cos\theta_1 d\cos\theta_2}= \frac{9}{4}\left[f_L \cos^2\theta_1 \cos^2\theta_2 + \frac{1}{4}(1-\ptrue) \sin^2\theta_1 \sin^2\theta_2 \right],
\end{eqnarray}
where $\theta_{i=1,2}$ are the angles between the \piz\ momentum 
and the direction opposite to that of the $B^0$ in the $\rho$ rest 
frame.

In this paper we present an update of previous \babar\ measurements
of the \Bztorhoprhom\ branching fraction, longitudinal polarization fraction,
\CP-violating parameters and measurement of the CKM angle $\alpha$, 
previously reported in Ref.~\cite{ref:us}. There are several theoretical 
predictions of the branching fraction and \ptrue\ in \Bztorhoprhom\ decays~\cite{kramer}.  The branching fraction of \Bztorhoprhom\ can be used 
to provide constraints on calculations of the branching fractions and 
charge asymmetries in $\B\to\pi\pi$ and $K\pi$ decays~\cite{ref:mishima}. 

The analysis reported here uses \lumi\ of data, which is significantly more 
than our previous branching fraction result (82 \ifb) or \ptrue\ and 
\CP-violating parameter 
 results (210 \ifb).  We have changed the selection requirements in order to
reduce background with only a modest reduction in signal efficiency. 
The new improved probability density-function (PDF) models
better account for the correlations between variables entering
the maximum likelihood (ML) fit, and result in a reduced systematic
uncertainty of the final results.

\section{THE \babar\ DETECTOR AND DATASET}
\label{sec:babar}
The data used in this analysis were collected with the \babar\ detector at the
\pep2\ asymmetric-energy \B\ Factory at SLAC. This represents a total 
integrated luminosity of \lumi\ taken at the $\Upsilon$(4S) 
resonance (on-peak), corresponding to a sample of \nbb. An additional 
\offpeaklumi\ of data, collected at approximately 40 \mev\ below 
the $\Upsilon$(4S) resonance (off-peak), were used to study 
background from non-resonant (continuum) \epem\to\qqbar events, 
where $\q = \u,\d,\s,\c$.

The detector is described in detail elsewhere~\cite{ref:babar}.
Surrounding the interaction point is a five double-sided layer silicon 
vertex tracker (SVT) which measures the impact parameters of charged 
particle tracks in both the plane transverse
to, and along the beam direction. A 40 layer drift chamber (DCH) surrounds the SVT and provides measurements
of the momenta for charged particles. Both the SVT and the DCH
 operate in the magnetic field of a 1.5 T solenoid. Charged hadron
identification is achieved through measurements of particle energy-loss
 in the tracking system and the Cherenkov angle obtained
from a detector of internally reflected Cherenkov light.
A CsI(Tl) electromagnetic calorimeter provides photon detection, electron
identification, and $\piz$ reconstruction.
Finally, the instrumented flux return of the magnet allows discrimination of muons from pions.
We use the GEANT4~\cite{ref:geant} software to simulate interactions of particles 
traversing the \babar\ detector.

\section{EVENT SELECTION}
\label{sec:Analysis}
We reconstruct \Bztorhoprhom\ candidates ($B_{\rm rec}$) from combinations of
two charged tracks and two \piz\ candidates. We require that both tracks have
particle identification information inconsistent with the electron, kaon, and
proton hypotheses. The \piz\ candidates are formed from pairs of photons, each 
of which has a measured energy greater than $50~\mev$. The reconstructed \piz\ mass 
must satisfy $0.10 < m_{\gamma\gamma} < 0.16~\gevcc$. The mass of the $\rho$ candidates
must satisfy $0.5 < \mv < 1.0~\gevcc$. 

Continuum events are the dominant background which is reduced by 
requiring that $|\cos(TB, TR)|$ be less than 0.8, where 
$|\cos(TB, TR)|$ is the absolute value of the cosine of the angle 
between the $B_{\rm rec}$ thrust axis and that of the rest of 
the event (ROE) calculated in the CM frame.  
To distinguish signal from continuum we use a neural network (\nno) combining the 
following eight discriminating variables.
\begin{itemize}
\item {The coefficients, $L_0, L_2$, where these are split
       into sums over the ROE for neutral and charged particles. 
       The coefficients are defined as:
       \begin{eqnarray}
       L_0 &=& \sum_{ROE} p_j\\
       L_2 &=& \sum_{ROE}p_j |\cos(\psi_j)|^2
       \end{eqnarray}
       where $p_j$ is the particle momentum and $\psi_j$ is the angle of 
       the particle direction relative to the thrust axis of the \B\-candidate in the center-of-mass (CM) frame.}  
\item{$|\cos(B, Z)|$, the absolute value of the cosine of the
      angle between the direction of the \B\ and $z$ axis in the CM frame,
      where the $z$ axis is along the electron beam direction.}  
\item{$|\cos(TB, TR)|$ as previously defined.} 
\item{$|\cos(TB, Z)|$, the absolute value of the cosine of the angle
      between the \B\ thrust and the $z$ axis.}
\item{the sum of the transverse momentum \pt\ of the ROE relative to 
      the $z$ axis.}  
\end{itemize}
Continuum backgrounds dominate near $|\coshel|=1$, and backgrounds from $B$ decays tend 
to concentrate at negative values of $\coshel$. We reduce these backgrounds with the
requirement $-0.90 < \coshel < 0.98$.

Signal events are identified kinematically using two variables, the
difference \DeltaE between the CM energy of the \B candidate $E_B$
and $\sqrt{s}/2$
\begin{equation}
\DeltaE = E_B - \sqrt{s}/2,
\end{equation}
and the beam-energy-substituted mass
\begin{equation}
 \mes = \sqrt{(s/2 + {\mathbf {p}}_i\cdot {\mathbf {p}}_B)^2/E_i^2- {\mathbf {p}}_B^2},
\end{equation}
where $\sqrt{s}$ is the total CM energy. The \B momentum
${\mathbf {p}_B}$ and four-momentum of the initial state $(E_i, {\mathbf
{p}_i})$ are measured in the laboratory frame. We accept candidates that
satisfy $5.25 < \mes <5.29~\gevcc$ and $-0.12<\DeltaE<0.15~\gev$. An
asymmetric \DeltaE\ selection is used in order to reduce background 
from higher-multiplicity \B decays.

When multiple \B\ candidates are formed, we select the one that minimizes the sum of $( m_{\gamma\gamma} - m_{\piz} )^2$
where $m_{\piz}$ is the \piz\ mass reported in~\cite{pdg}.
In 0.3\% the same \piz\ mesons are used by multiple \B\ candiadtes.  
We randomly select the candidate entering the fit for such events.

In order to study the time-dependent decay-rate asymmetry 
one needs to measure the proper-time difference \deltat\ between 
the two \B\ decays in the event, and to determine the flavor 
of the other \B\ meson ($B_{\rm tag}$).
We calculate \deltat\ from the measured separation
\deltaz between the $B_{\rm rec}$ and $B_{\rm tag}$ decay
vertices~\cite{prdsin2b}. We determine the $B_{\rm rec}$ vertex
from the two charged-pion tracks in its decay. The $B_{\rm tag}$
decay vertex is obtained by fitting the other tracks in the event,
with constraints from the $B_{\rm rec}$ momentum and the
beam-spot location. The RMS resolution on
$\deltat$ is 1.1 \ps. We only use events that satisfy 
$|\deltat|<15 \, \ps$ and have an error on \deltat of less than $2.5 \, \ps$.
The flavor of the $B_{\rm tag}$ meson is determined with a multivariate
technique~\cite{pipiBabar}.  The performance of this algorithm is summarised in
Table~\ref{tbl:taggingeff} for the seven mutually-exclusive tag categories.  
Events are categorised according to decreasing signal purity and increasing 
mistag probability $\omega$.
The categories assigned correspond to events with leptons, kaons and
pions in the decay products of $B_{\rm tag}$.  The \notag category of events are 
dominated by continuum background, have a mistag probability of 50\%, 
and are not used in this analysis.
\begin{table}[!ht]
\caption{Tagging efficiency $\epsilon_c$, average mistag fraction  $\omega_c$, and mistag 
fraction difference $\Delta \omega_c$ between \Bz\ and \Bzb\ tagged events for \Bztorhoprhom\ events in each tagging category.}
\label{tbl:taggingeff}
\begin{center}
\begin{tabular}{lccc} \hline
Category $(c)$ & $\epsilon_{c}$ & $\omega_{c}$ & $\Delta\omega_{c}$ \\ \hline
\lepton    & 0.080 $\pm$ 0.001 & 0.032 $\pm$ 0.004 & 0.002 $\pm$ 0.008 \\
\kaon\ 1   & 0.114 $\pm$ 0.001 & 0.053 $\pm$ 0.005 & -0.012 $\pm$ 0.009 \\
\kaon\ 2   & 0.176 $\pm$ 0.002 & 0.152 $\pm$ 0.005 & -0.015 $\pm$ 0.008 \\
\kaonpion  & 0.135 $\pm$ 0.002 & 0.233 $\pm$ 0.006 & -0.018 $\pm$ 0.010 \\
\pion      & 0.137 $\pm$ 0.002 & 0.327 $\pm$ 0.007 & 0.059 $\pm$ 0.010 \\
\other     & 0.095 $\pm$ 0.001 & 0.412 $\pm$ 0.008 & 0.042 $\pm$ 0.001 \\
\notag     & 0.266 $\pm$ 0.002 & 0.500 $\pm$ 0.000 & 0.000 $\pm$ 0.000 \\
 \hline
\end{tabular}
\end{center}
\end{table}

Mis-reconstructed signal candidates or self-cross-feed (SCF) signal may pass the selection requirements even 
if one or more of the pions assigned to the $\rho^+\rho^-$ state belongs to 
the other \B\ in the event. These SCF 
candidates constitute 50.7\% (27.9\%) of the accepted longitudinally (transversely) polarised signal. 
The majority of SCF events have both charged pions from the 
$\rho^+\rho^-$ final state, and unbiased \CP\ information. 
These correct track SCF events are denoted by CT SCF.  There 
is a SCF component (13.8\% of the signal) where at least one track in $B_{\rm rec}$ is from the ROE.  
These wrong track (WT) events have biased \CP information, and are treated 
separately for the \CP result. The PDF describing WT events is used only to 
determine the signal yield 
and polarization. A systematic error is assigned to the \CP 
results from this type of signal event.  The total selection efficiency
for longitudinally (transversely) polarised signal is 7.7\% (10.5\%).

\section{MAXIMUM LIKELIHOOD FIT}
We obtain a sample of \ndata\ events that enter an unbinned extended ML fit.  
These events are dominated by backgrounds from \qqbar\ ($81.4\%$)
and \BB\ ($16.6$\%) events. The remaining 2\% of events are signal. 
We distinguish between the following components in the fit: 
 (i) correctly reconstructed signal;
 (ii) SCF signal, split into CT and WT parts;
 (iii) charm $\Bpm$ background ($b\to c$);
 (iv) charm $\Bz$ background ($b\to c$);
 (v) charmless $\Bz$ backgrounds;
 (vi) charmless $\Bpm$ backgrounds; and
 (vii) continuum background. 
The dominant \B\ backgrounds come from categories (iii) and (iv).
The dominant charmless backgrounds are summarised in Table~\ref{table:charmless}.
The branching fractions of $\Bch\to (a_1 \pi)^\pm$ decays have been estimated using
the measurements of $\Bz\to a_1^\pm \pi^\mp$~\cite{aonepi}.
\begin{table}[h!]
\caption{Dominant charmless backgrounds with assumed or measured branching 
fraction and the number of selected events in the data sample.\label{table:charmless}}
\begin{center}
\begin{tabular}{lcc} \hline
Decay                       & Branching fraction ($10^{-6}$) & Number of events \\ \hline
$a_1^\pm \pi^\mp$           & $39.7\pm 3.7$~\cite{aonepi}                          & $94 \pm 9$  \\
$\rho^0\rho^+$ (long)       & $19.1 \pm 3.5$~\cite{rhorho0}                        & $70 \pm 13$ \\
$a_1^+ \pi^0$               & $20\pm 20$                                           & $55 \pm 55$ \\
$a_1^0 \pi^+$               & $20\pm 20$                                           & $45 \pm 45$ \\
$a_1^\pm\rho^\mp$ (long)    & $24 \pm 2.5$~\cite{rhopi}                            & $40 \pm 4$  \\
$\rho^\pm\pi^\mp$           & $30 \pm 30$~\cite{aonerho}                           & $40 \pm 40$ \\
 \hline
\end{tabular}
\end{center}
\end{table}

We allow the charm and the non-resonant $\Bz\to\rho^+\pi^-\pi^0$ yields to vary in the fit to data. 
The remaining background yields are fixed to their expected values determined from measured branching
fractions where available~\cite{ref:hfag}.

The likelihood function incorporates the following previously defined eight variables to distinguish signal from background: 
\mes, \DeltaE, \deltat, \nno, and the \mv, and \coshel\ values of the two $\rho$ mesons.
For each of the aforementioned components we construct a PDF that is the product of  
one-dimensional PDFs for each of the variables.  The PDFs for all of the components are
combined to give the likelihood function used in the fit.
Table~\ref{table:pdfs} summarises the PDF shapes used for different components of the signal.   

\begin{table*}[!ht]
\caption{PDFs used to parameterize signal distributions. The abbreviations used are as follows:
\Long $-$ {\it longitudinal signal Monte Carlo (MC) simulated data};
\Tran $-$ {\it transverse signal MC simulated data};
\true $-$ {\it true signal (or in the case of background, true $\rho$ meson)};
\SCF $-$ {\it SCF signal};
pN $-$ {\it polynomial of order N};
G $-$ {\it a Gaussian distribution};
CB $-$ {\it a Gaussian with an exponential tail which takes the form 
  of~\cite{crystalball}};
NP $-$ {\it a non-parametric PDF defined as smoothed histogram of MC 
  simulated data};
RBW $-$ {\it a relativistic Breit-Wigner};
SigDT $-$ {\it a distribution of the form of Eq.~\ref{eq:sigdt}, convoluted with a triple Gaussian resolution function.}
Where indicated, the PDF denoted by pN acc. refers to the physical distribution multiplied by a 
polynomial acceptance function of order N. 
}\label{table:pdfs}
\small
\begin{center}
\begin{tabular}{c|cccccccc} \hline

Component                     & \mes & \de  & \nno & \dt   & \multicolumn{2}{c}{$\cos\theta_{\rho}$} & \multicolumn{2}{c}{$m_{\rho}$}\\ 
                              &      &      &      &       &  True         &         Fake     & True         &         Fake     \\ \hline  

true (\Long)                  & CB   & CB+G & NP   & SigDT & p4 acc.                   & N.A.                       & RBW              & N.A.\\
true (\Tran)                  & CB+G & CB+G & NP   & SigDT & p4 acc.                   & N.A.                       & RBW              & N.A.\\
CT \SCF (\Long) (TT)          & CB   & p2+G & NP   & SigDT & p6 acc.                   & NP                         & RBW              & p3 \\
CT \SCF (\Long) (TF/FT)       & CB+G & p2   & NP   & SigDT & p6 acc.                   & NP                         & RBW              & p3 \\
CT \SCF (\Long) (FF)          & CB+G & p1   & NP   & SigDT & p6 acc.                   & NP                         & RBW              & p3 \\
WT \SCF (\Long) (TF/FT)       & CB+G & p2   & NP   & SigDT & p6 acc.                   & NP                         & RBW              & p3 \\
WT \SCF (\Long) (FF)          & CB+G & p1   & NP   & SigDT & p6 acc.                   & NP                         & RBW              & p3 \\
\SCF (\Tran)                  & CB+G & GG   & NP   & SigDT & NP                        & NP                         & NP               & NP \\\hline
\end{tabular}
\end{center}
\end{table*}

The \B\ background \de\ distributions are described by $3^{rd}$ order polynomials, except for non-resonant 
$\Bz\to\rho^+\pi^-\piz$ which uses a non-parametric (NP) PDF.  The $m_\rho$ distributions 
for true $\rho$ mesons are parametersised using a 
relativistic Breit-Wigner, and the fake $m_\rho$ (combinatorial 
$\pi^\pm\piz$) distribution is described using  $3^{rd}$ order polynomials. The \mes\ distribution of $b\to c$ backgrounds 
is described using a phase-space-motivated distribution~\cite{argusfunction}.  All remaining background 
shapes are described using NP PDFs.  

The continuum distribution for \mes\ is described by a phase-space-motivated 
distribution~\cite{argusfunction}.  The \de\ and \nno\ shapes are 
modeled with  $3^{rd}$ and $4^{th}$ order polynomials, respectively.  The 
parameters of the \mes, \de\ and \nno\ shapes are allowed to vary 
in the fit to data. The continuum \mv\ distribution is described using a 
relativistic Breit-Wigner and a $3^{rd}$ order polynomial PDF for true and fake $\rho$ contributions, respectively.  The \coshel\ 
distribution is described by a  $3^{rd}$ order polynomial.  The parameters of the \mv\ and \coshel\ distributions 
are obtained from a fit to the off-peak data.

The signal decay-rate distribution for both polarizations
$f_+ (f_-)$ for $B_{\rm tag}$= \Bz (\Bzb) is given by
\begin{eqnarray}
f_{\pm}(\deltat) = \frac{e^{-\left|\deltat\right|/\tau}}{4\tau} [1 \pm S\sin(\deltamd\deltat) \mp \C\cos(\deltamd\deltat)]\label{eq:sigdt}
\end{eqnarray}
where $\tau$ is the mean \Bz lifetime, \deltamd is the \BzBzb\
mixing frequency, and $S$ = \slong\ or \stran\ and $C$ = \clong\ or
\ctran\ are the \CP-asymmetry parameters for the longitudinally and
transversely polarized signal. The parameters $S$ and $C$ describe 
\B-mixing induced and direct \CP violation, respectively. 
$S$ and $C$ for the longitudinally
polarized WT signal are set to zero. The \deltat PDF
 takes into account incorrect tags and is convolved with
the resolution function described below. Since \ptrue\ is approximately $1$, the fit has no
sensitivity to either \stran\ or \ctran. We set these parameters
to zero and we vary them in the evaluation of systematic
uncertainties.

The signal \deltat\ resolution function consists of three
Gaussians ($\sim 85\%$ core, $\sim 14\%$ tail, $\sim 1\%$ outliers),
and takes into account the per-event error on $\deltat$ from the 
vertex fit. The resolution function parameters are obtained from 
a large sample of fully reconstructed hadronic \B decays~\cite{prdsin2b}. For WT 
SCF we replace the \B-meson lifetime by an effective lifetime obtained from 
Monte Carlo (MC) simulation to account for the difference in the resolution. The nominal \deltat\
distribution for the \B backgrounds is a NP PDF
representation of the MC samples; in the study of systematic
errors we replace this model with the one used for signal. The
resolution for continuum background is described by the sum of three Gaussian
distributions whose parameters are determined from data.

\section{RESULTS}
\label{sec:Physics}

From the extended maximum likelihood fit described above,
we obtain the following results
\begin{eqnarray}
N({\rm signal}) &=& \correctedsignalyield,\nonumber \\
\ptrue &=& \correctedfl,\nonumber \\
\slong &=& \correctedslong,\nonumber \\
\clong &=& \correctedclong,\nonumber 
\end{eqnarray}
after correction for a $+28$ event fit bias on the signal yield and
a correction for a $-0.007$ fit bias on \ptrue.
We discuss the origin of these fit biases in Section~\ref{sec:Systematics}.
The $\Bz\to\rho^\pm\pi^\mp\piz$ background yield obtained from the fit is $9.2 \pm 53.6$ events.
Figure~\ref{fig:plots} shows
distributions of \mes, \DeltaE, \coshel\ and \mv\ for the highest purity 
tagged events with a loose 
requirement on \nno. The plot of \mes\ contains 15.6\% of the 
signal and 1.1\% of the background. For the other plots 
there is an added constraint that $\mes > 5.27 \gevcc$; these 
requirements retain 13.9\% of the signal and 0.4\% of the background.  
Figure~\ref{fig:dtplots} shows
the \deltat\ distribution for \Bz and \Bzb
tagged events. The time-dependent decay-rate asymmetry 
\begin{equation}
a_{\CP}(\deltat) = \frac{ N (\deltat) - \overline{N}(\deltat) } { N (\deltat) + \overline{N}(\deltat) }
\end{equation}
is also shown, where $N$ $(\overline{N})$ is the decay-rate for \Bz(\Bzb) tagged events.

\begin{figure}[!ht]
\begin{center}
\resizebox{12cm}{!}{
%
%
 \includegraphics{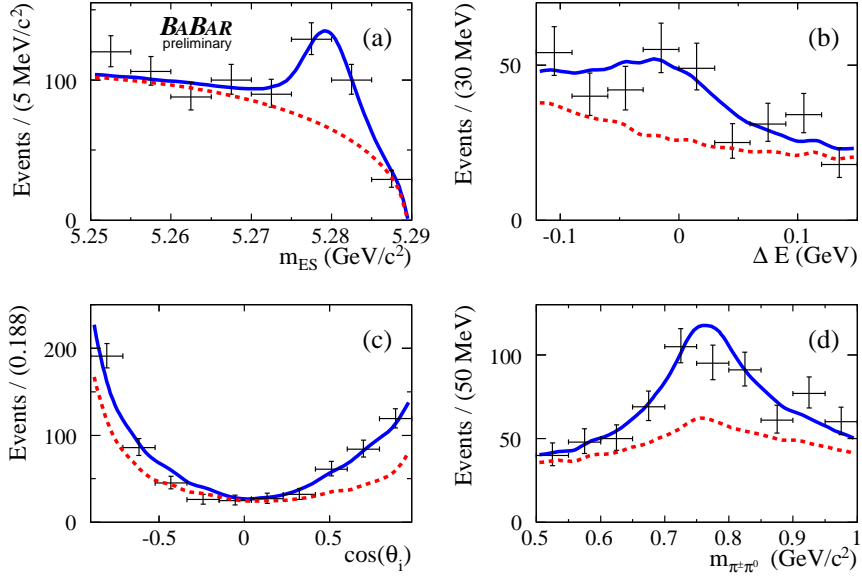}
}

\caption{The distributions for the highest purity tagged events for
the variables \mes\ (a),  \DeltaE\ (b),  cosine of the $\rho$ helicity
angle (c) and  \mv\ (d).  The dashed lines are the sum of backgrounds,  
and the solid lines are the full PDF.
} \label{fig:plots}
\end{center}
\end{figure}

\begin{figure}[!ht]
\begin{center}
\resizebox{12cm}{!}{
 \includegraphics{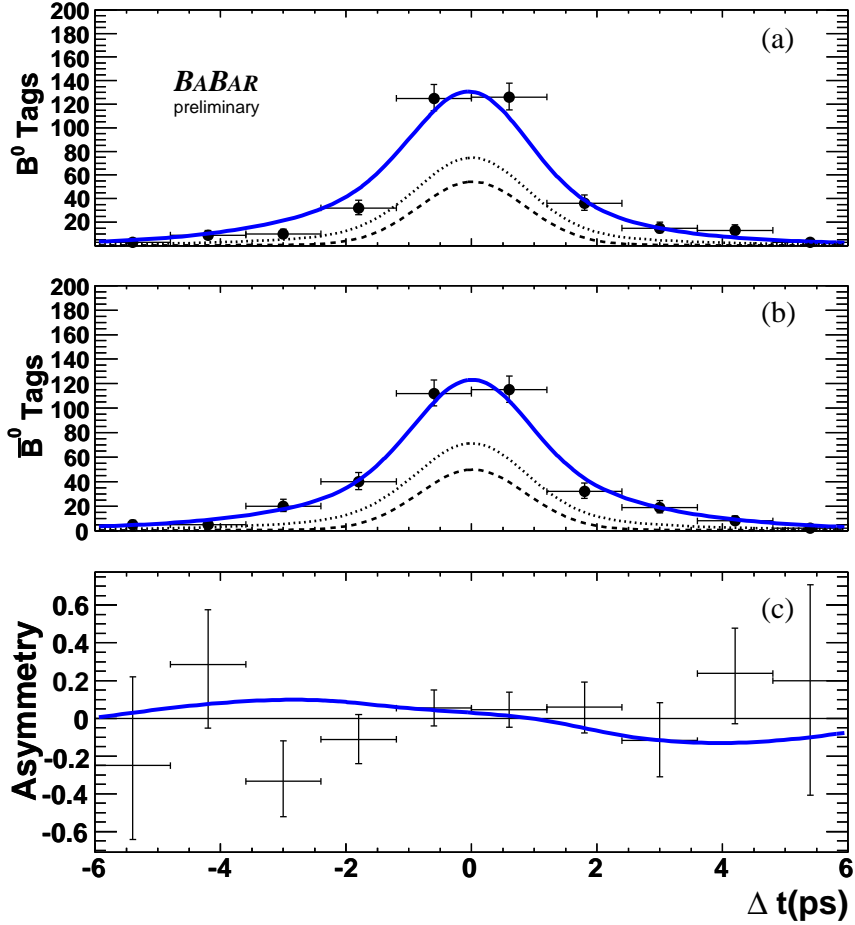}
}
 \caption{The \deltat\ distribution for a sample of events enriched in signal for 
  (a) $\Bz$ and (b) $\Bzb$ tagged events. The dotted lines are the sum of backgrounds,
  the dashed lines are the \qqbar\ background and the solid lines are the sum of signal 
  and backgrounds.  The time-dependent \CP asymmetry (see text) is shown in (c), where 
  the curve is the measured asymmetry.} \label{fig:dtplots}
\end{center}
\end{figure}

\section{SYSTEMATIC ERROR STUDIES}
\label{sec:Systematics}

Table~\ref{tbl:systsummary} lists the possible sources of systematic uncertainties 
on the values of the signal yield, \ptrue, \slong\ and \clong\ that have been studied.
The systematic uncertainties, listed in Table~\ref{tbl:systsummary}, are briefly 
discussed in the following.
\begin{itemize} 
\item The uncertainty from PDF parameterisation is obtained by varying PDF shape parameters by 
      $\pm 1 \sigma$, in turn.  The deviations obtained are added in quadrature
      to give the quoted uncertainty from the PDF parameterisation.
\item The systematic uncertainty on the yield from the fraction of \SCF\ events is obtained 
      taking the difference between the nominal result and the number of events fitted (after 
      correcting for the fraction of \SCF\ events in MC simulated data) when using the true 
      signal PDFs to extract the yield.  The uncertainty on the other signal observables comes 
      from varying the fraction of \SCF\ events by 5\% per \piz\ (10\% total), which is twice 
      the observed data/MC difference seen in our control sample of $\Bz\to D^-\rho^+$ events.  
\item The uncertainty on the \mes\ and \de\ widths is obtained from the
      observed shifts relative to our nominal result, when allowing these parameters to vary independently in the fit to data.  
\item We vary the \B\ background normalisation within expectations for each background in turn.  The
      deviations obtained when doing this are added in quadrature
      to give the quoted uncertainty from this source.
\item As the branching fractions of some of the \B\ backgrounds are not well known, we assign an additional 
      uncertainty coming from the maximum shifts obtained when allowing each of the fixed backgrounds
      to vary in turn in the fit to data.  
\item Additional uncertainties on the \CP results come from possible \CP violation in the \B\ background.  
      We use existing experimental constraints where possible, 
      otherwise we allow for a \CP asymmetry up to 10\% in \B\ decays 
      to final states with charm, and up to 50\% in \B\ decays to charmless 
      final states.  
\item The physics parameters $\tau=1.530\pm 0.009$ \ps\ and $\deltamd = 0.507 \pm 0.005$ $\hbar \ps$~\cite{pdg} 
      are varied within the quoted uncertainty.  
\item The tagging and mistag fractions for signal and the \B\ backgrounds are corrected for data/MC 
      differences observed in samples of fully reconstructed hadronic \B\ decays.  Each of the tagging 
      and mistag parameters is varied in turn by the uncertainty from the correction.
      The deviations obtained when doing this are added in quadrature
      to give the quoted uncertainty from this source.
\item Allowing for possible \CP\ violation in the transverse polarization, and in the WT 
      longitudinally polarised signal \SCF\ events results in additional uncertainties on signal observables.
      We vary $S$ and $C$ by $\pm 0.5$ ($\pm 1.0$) for the transverse polarisation (WT \SCF).
\item Possible \CP\ violation from interference in doubly Cabibbo-suppressed decays (DCSD) on the tag side of the 
      event~\cite{ref:dcsd} contribute to systematic uncertainties on \slong\ and \clong.  
\item We estimate the systematic error on our results coming from neglecting the interference between 
      \Bztorhoprhom\ and other $4\pi$ final states: $\B \to a_1\pi$, $\rho \pi\pi^0$ and $\B\to\pi\pi\piz\piz$. 
      Strong phases are varied between $-180$ and $180$ degrees, and the \CP\ content of the interfering 
      amplitudes are varied between zero and maximum using uniform prior distributions,
      and the RMS deviation of the parameters from nominal is taken as the 
      systematic error.  We assume that the amplitudes for $\B \to \rho \pi\pi^0$ and 
      $\pi\pi\piz\piz$ are equal to the amplitude of $\B \to a_1\pi$ when calculating this
      systematic uncertainty.
\item As the PDFs used in the ML fit do not account for all of the correlations between discriminating 
      variables used in the fit, the results have a small bias.  We calculate the fit bias on the signal 
      observables from ensembles of experiments obtained from samples of signal and charmless 
      \B\ background MC simulated events combined with charm and \qqbar\ background events simulated 
      from the PDFs.  The 28 event bias on signal yield and 0.007 bias on \ptrue\ is corrected, and 
      100\% of the calculated fit bias is assigned as a systematic uncertainty on 
      all signal observables.  We do not observe a significant bias on \slong\ and \clong.
\item Small imperfections in the knowledge of the geometry of the SVT over time can affect the 
      measurement of \slong\ and \clong.  We vary the alignment according to the results obtained from the 
      study of $\epem \to e^+e^-, \mu^+\mu^-$ events in order to estimate the magnitude of this systematic error
      on our \CP\ results.
\end{itemize}
\begin{table}[!ht]
\caption{Summary of additive systematic uncertainty contributions.}\label{tbl:systsummary}
\begin{center}
\begin{tabular}{l|cccc}
Contribution                 & $\sigma(N_{signal})$ & $\sigma(\ptrue)$ & $\sigma(\slong)$ & $\sigma(\clong)$  \\ \hline\\
PDF parameterisation         & $^{+16.7}_{-30.2}$ & $^{+0.0082}_{-0.0064}$ & $^{+0.0149}_{-0.0425}$ & $^{+0.0300}_{-0.0306}$ \\[6pt]
\SCF\ fraction               & $84.0$             & $^{+0.0007}_{-0.0011}$ & $^{+0.00235}_{-0.00355}$ & $^{+0.0070}_{-0.00683}$ \\[6pt]
\mes\ and \de\ width         & $22.9$             & $0.005$                & $0.011$                & $0.012$ \\[6pt]
\B\ background normalisation & $^{+16.0}_{-17.2}$ & $^{+0.0033}_{-0.0038}$ & $^{+0.0096}_{-0.0115}$ & $^{+0.0024}_{-0.0015}$ \\[6pt]
floating \B\ backgrounds     & $33.6$             & $0.004$                & $0.033$                & $0.006$ \\[6pt]
\CPV\ in \B\ background      & $^{+3.3}_{-2.0}$   & $^{+0.0006}_{-0.0016}$ & $^{+0.0059}_{-0.0214}$ & $^{+0.0118}_{-0.0115}$ \\[6pt]
$\tau$                       & $^{+0.1}_{-0.4}$   & $^{+0.0000}_{-0.0002}$ & $^{+0.0002}_{-0.0008}$ & $0.0007$ \\[6pt]
\dm                          & $^{+0.0}_{-0.2}$   & $^{+0.0000}_{-0.0002}$ & $^{+0.0014}_{-0.0020}$ & $^{+0.0018}_{-0.0012}$ \\[6pt]
tagging and dilution         & $^{+2.6}_{-8.1}$   & $^{+0.0029}_{-0.0021}$ & $^{+0.0016}_{-0.0053}$ & $^{+0.0068}_{-0.0054}$ \\[6pt]
transverse polarisation \CPV & $^{+0.0}_{-8.3}$    & $^{+0.0057}_{-0.0000}$ & $^{+0.0125}_{-0.0152}$ & $^{+0.0095}_{-0.0110}$ \\[6pt]
WT \SCF\ \CPV       & $^{+0.2}_{-1.1}$   & $^{+0.0000}_{-0.0003}$ & $^{+0.0051}_{-0.0065}$ & $^{+0.0116}_{-0.0113}$ \\[6pt]
DCSD decays                  & $-$                & $-$                    & $0.012$                & $0.037$ \\[6pt]
Interference                 & $14.8$            & $0.0036$               & $0.023$                & $0.022$ \\[6pt]
Fit Bias                     & $28$               & $0.007$                & $0.002$                & $0.022$ \\[6pt]
SVT Alignment                & $-$                & $-$                    & $0.0100$               & $0.0055$ \\[6pt] \hline \\
Total                        & $^{+97}_{-101}$ & $^{+0.015}_{-0.013}$   & $^{+0.05}_{-0.07}$   & $\pm 0.06$ \\[6pt] \hline
\end{tabular}
\end{center}
\end{table}

The branching fraction has multiplicative systematic uncertainties from 
the reconstruction of \piz\ mesons in the detector (6\%), uncertainties
in the reconstruction of charged particles (0.8\%), and the discrimination
of $\pi^\pm$ from other types of charged particles (1\%).  In addition to these 
uncertainties, there is a 1.1\% uncertainty on the number of \bb\ pairs in the
data sample.  Statistical uncertainties arising from the MC samples 
used in this analysis are negligible.

\clearpage

\section{FINAL RESULTS}
\label{sec:ResultsSummary}

Our results are \vspace{-0.9cm}
\begin{center}
\begin{eqnarray}
{\cal B}(\Bztorhoprhom) &=& \measuredbf,\nonumber \\
\ptrue &=& \measuredfl,\nonumber \\
\slong &=& \measuredslong,\nonumber \\
\clong &=& \measuredclong.\nonumber 
\end{eqnarray}
\end{center}\vspace{-0.2cm}
where the correlation between {\slong} and \clong\ is
\sccorrelation.  This measurement corresponds to 
$\alpha_{\mathrm{eff}} = (95.5^{+6.9}_{-6.2})^\circ$,
where $\slong = \sqrt{1-\clong^2}\sin 2 \aeff$.
The measured branching fraction, polarization, and
\CP parameters are in agreement with our earlier
publication~\cite{ref:us}, with significantly improved precision.

We constrain the CKM angle $\alpha$ from an isospin 
analysis~\cite{grossmanquinn} of $B \to \rho\rho$.  The 
inputs to the isospin analysis are the amplitudes of 
the \CP-even longitudinal polarization of the $\rho\rho$ final state, as 
well as the measured values of \slong\ and \clong\
for \Bztorhoprhom.  We use the following numerical inputs in the isospin analysis:
\begin{itemize}
 \item The average of the measurements of ${\cal B}(\Bztorhoprhom)$ and \ptrue, presented here, and those
       reported in Ref.~\cite{bellerhoprhom}.
 \item The combined branching fraction and \ptrue\ for $\B\to\rhop\rhoz$ from Ref.~\cite{rhorho0}. 
 \item The central value corresponding to the upper limit of ${ \cal B}(\B\to\rhoz\rhoz)$ from Ref.~\cite{PRLrho0rho0}.
 \item \slong\ and \clong\ presented here. 
\end{itemize}
We ignore possible $I=1$ amplitudes~\cite{falk} and electroweak penguins in this paper.

To interpret our results in terms of a constraint on $\alpha$ from
the isospin relations,  we construct a $\chi^2$ that includes
the measured quantities expressed as the lengths of the sides of
the isospin triangles and we determine the minimum $\chi^2_0$.
We have adopted a simulated experiment technique to compute the confidence 
level (CL) on $\alpha$; our method is similar to
the approach proposed in Ref.~\cite{FC98}. 
For each value of $\alpha$, scanned between
$0$ and $180^\circ$, we determine the
difference $\Delta \chi^2_{{\rm DATA}}(\alpha)$  between  the minimum
of $\chi^2(\alpha)$ and $\chi^2_0$. We then generate MC experiments
around the central values obtained from the fit to data with
the given value of $\alpha$ and
we apply the same procedure. The fraction of these experiments in
which  $\Delta \chi^2_{{\rm MC}}(\alpha)$ is smaller than
$\Delta \chi^2_{{\rm DATA}}(\alpha)$ 
is interpreted as the CL on $\alpha$.
Figure~\ref{fig:alpha} shows $1-{\rm CL}$ for
$\alpha$ obtained from this method.
Selecting the solution closest to the CKM
combined fit average~\cite{ref:ckmbestfit,ref:utfit} we find $\alpha =
\measuredalpha$, where the error is
dominated by $|\alpha_{\mathrm{eff}} - \alpha|$ which is 
$\measureddeltaalpha$ at $68\%$ CL. 
The constraint obtained on $\alpha$ is less precise than the previous
result in Ref.~\cite{ref:us} for the following reasons: (i) the 
improved world average branching fraction for
$\Bp \to \rho^+\rho^0$ has come down, and (ii) the results of the latest
search for \Bztorhozrhoz\ show evidence for a signal.  Both of
these factors lead to an increase in the the penguin contribution
to the total uncertainty on $\alpha$ when using an isospin analysis.

\begin{figure}[!ht]
 \begin{center}
 \resizebox{12cm}{!}{
 \includegraphics[width=0.48\textwidth]{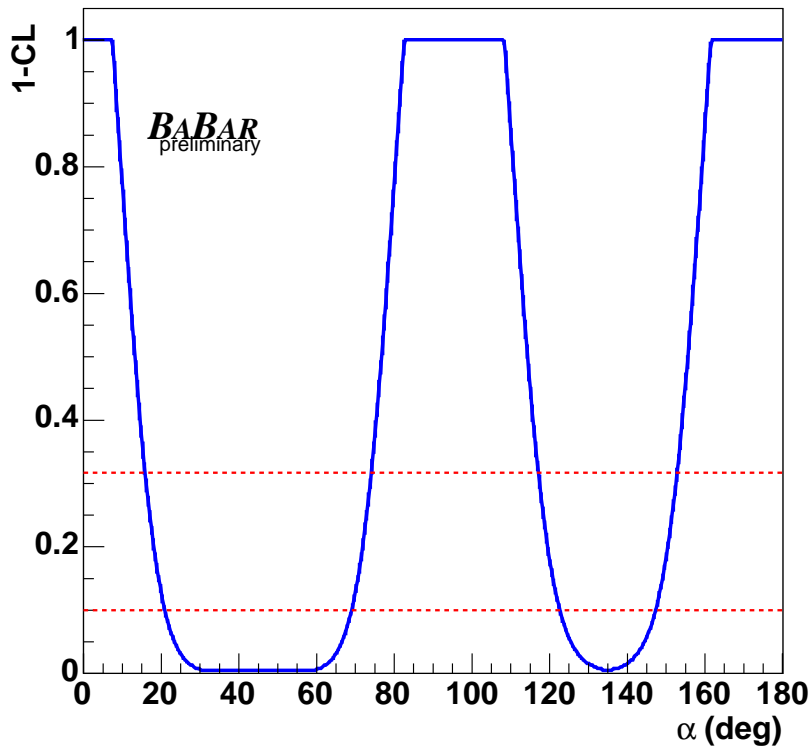}
} 
\caption{ Confidence level on $\alpha$ obtained from the isospin analysis 
with the statistical method described in~\cite{ref:ckmbestfit}. 
The dashed lines correspond to the 68\% (top) and 90\% (bottom) 
CL intervals. } \label{fig:alpha}
 \end{center}
\end{figure}

It is possible to calculate a model-dependent constraint on $\alpha$ using 
\su{3} as detailed in Ref.~\cite{beneke}.  This model relates the penguin 
amplitude in $B^+ \to K^{*0}\rho^+$ to the penguin amplitude in 
\Bztorhoprhom, and can be used to obtain a more precise, but model 
dependent constraint on $\alpha$.  

\section{SUMMARY}
\label{sec:Summary}

In summary we have improved the measurement of the branching fraction, \ptrue, 
and \CP-violating parameters \slong\ and \clong\ in \Bztorhoprhom\ using a 
data-sample of \nbb. We do not observe mixing-induced or direct 
\CP violation. We derive a model-independent measurement of the 
CKM angle $\alpha$.  The results presented in this paper are consistent with
previous results presented by \babar~\cite{ref:us} and Belle~\cite{bellerhoprhom}.
\section{ACKNOWLEDGMENTS}
\label{sec:Acknowledgments}

We are grateful for the 
extraordinary contributions of our \pep2\ colleagues in
achieving the excellent luminosity and machine conditions
that have made this work possible.
The success of this project also relies critically on the 
expertise and dedication of the computing organizations that 
support \babar.
The collaborating institutions wish to thank 
SLAC for its support and the kind hospitality extended to them. 
This work is supported by the
US Department of Energy
and National Science Foundation, the
Natural Sciences and Engineering Research Council (Canada),
Institute of High Energy Physics (China), the
Commissariat \`a l'Energie Atomique and
Institut National de Physique Nucl\'eaire et de Physique des Particules
(France), the
Bundesministerium f\"ur Bildung und Forschung and
Deutsche Forschungsgemeinschaft
(Germany), the
Istituto Nazionale di Fisica Nucleare (Italy),
the Foundation for Fundamental Research on Matter (The Netherlands),
the Research Council of Norway, the
Ministry of Science and Technology of the Russian Federation, and the
Particle Physics and Astronomy Research Council (United Kingdom). 
Individuals have received support from 
the Marie-Curie IEF program (European Union) and
the A. P. Sloan Foundation.


\begin{thebibliography}{99}
\bibitem{crystalball}
The definition used for a Gaussian with exponential tail is given by \newline
$CB(m\,;m_0,\sigma,\alpha,n) = \frac{1}{N}\,\cdot\,\exp\left(-(m-m_0)^2/(2\sigma^2)\right) ,\; m > m_0 - \alpha\sigma$ \newline
and $\frac{1}{N}\,\cdot\,\dfrac{(n/\alpha)^n\,\exp(-\alpha^2/2)} {\left((m_0 - m)/\sigma + n/\alpha - \alpha\right)^n} ,\; m \leq m_0 - \alpha\sigma$.
\end{thebibliography}
\end{document}